\documentclass[pra,twocolumn,amsmath,amssymb,groupaddress,eqsecnum,letterpape]{revtex4-2}
\usepackage{graphicx,amsmath,relsize,epstopdf,color,mathtools,bm,newtxtext,newtxmath,braket,rotating,comment}
\usepackage[hyphenbreaks]{breakurl}
\usepackage[colorlinks=true,linkcolor=blue,citecolor=blue,urlcolor =blue]{hyperref}
\usepackage[normalem]{ulem}
\usepackage[table,xcdraw]{xcolor}

\newcommand{\Tr}{\mathop{\mathrm{Tr}} \nolimits}
\newcommand{\Var}{\mathop{\mathrm{Var}}\nolimits}

\usepackage{soul}

\begin{document}

\title{Robust quantum metrology with random Majorana constellations}

\author{Aaron~Z.~Goldberg}
\affiliation{National Research Council of Canada, 100 Sussex Drive, Ottawa, Ontario K1N 5A2, Canada}

\author{Jose~R.~Herv\'as}
\affiliation{Departamento de \'Optica, Facultad de F\'{\i}sica,
Universidad Complutense, 28040 Madrid, Spain}

\author{Angel~S.~Sanz}
\affiliation{Departamento de \'Optica, Facultad de F\'{\i}sica,
Universidad Complutense, 28040 Madrid, Spain}

\author{Andrei~B.~Klimov}
\affiliation{Departamento de Física, Universidad de Guadalajara, 
44420~Guadalajara, Jalisco, Mexico}

\author{Jaroslav~\v{R}eh\'a\v{c}ek}
\affiliation{Department of Optics, Palack\'{y} University, 17. listopadu 12, 771 46 Olomouc, Czech Republic}

\author{Zden\v{e}k~Hradil}
\affiliation{Department of Optics, Palack\'{y} University, 17. listopadu 12, 771 46 Olomouc, Czech Republic}

\author{Markus~Hiekkam\"aki}
\affiliation{Physics Unit, Photonics Laboratory, Tampere University, 33720 Tampere, Finland}

\author{Matias~Eriksson}
\affiliation{Physics Unit, Photonics Laboratory, Tampere University, 33720 Tampere, Finland}

\author{Robert~Fickler}
\affiliation{Physics Unit, Photonics Laboratory, Tampere University, 33720 Tampere, Finland}

\author{Gerd~Leuchs}
\affiliation{Max-Planck-Institut f\"{u}r die Physik des Lichts, 91058 Erlangen, Germany}

\author{Luis~L. S\'{a}nchez-Soto$^\ast$}
\affiliation{Max-Planck-Institut f\"{u}r die Physik des Lichts, 91058 Erlangen, Germany}

\begin{abstract}
Even the most classical states are still governed by quantum theory. A number of physical systems can be described by their Majorana constellations of points on the surface of a sphere, where concentrated constellations and highly symmetric distributions correspond to the least and most quantum states, respectively. If these points are chosen randomly, how quantum will the resultant state be, on average? We explore this simple conceptual question in detail, investigating the quantum properties of the resulting random states. We find these states to be far from the norm, even in the large-number-of-particles limit, where classical intuition often replaces quantum properties, making random Majorana constellations peculiar and intriguing. Moreover, we study their usefulness in the context of rotation sensing and find numerical evidence of their robustness against dephasing and particle loss. We realize these states experimentally using light's orbital angular momentum degree of freedom and implement arbitrary unitaries with a multiplane light conversion setup to demonstrate the rotation sensing. Our findings open up new possibilities for quantum-enhanced metrology.
\bigskip
\bigskip

$^\ast$ Author to whom any correspondence should be addressed. e-mail: \url{lsanchez@ucm.es} 
\end{abstract}

\maketitle

\section{Introduction}
 
Random matrices sampled from an ensemble with a specific symmetry were introduced by Wigner~\cite{Wigner:1955wg,Wigner:1967ve} and Dyson~\cite{Dyson:1962uy,Dyson:1962wh,Dyson:1962vk} to describe spectral properties of quantum many-body systems, such as atomic nuclei~\cite{Brody:1981ur}. Since then, random matrix theory~\cite{Mehta:2004wq,Tao:2012vt,Livan:2018us} has found applications in fields as diverse as black holes~\cite{Page:1993ur,Page:1993th,Sen:1996wo,Renner:2021vj} and gravity~\cite{Francesco:1995ua}, quantum chaos~\cite{Bohigas:1984wu}, transport in disordered systems~\cite{Mello:1990wc,Beenakker:1997ui}, spin glasses~\cite{Cugliandolo:1995uw}, neural networks~\cite{Sompolinsky:1988vy,Wainrib:2013vs}, and even finance~\cite{Bouchaud:2000up}. Quantum information is definitely one of the most recent applications, and a very natural one, too~\cite{Collins:2015uo,Russell:2017wh,Boixo:2018un}. 

Random quantum states can be seen as arising from the time evolution of arbitrary initial states of quantum analogues of classically chaotic systems~\cite{Haake:1991tm}. Furthermore, these states emerge when not much is known about a state and one wants to ask about its generic properties, characteristic of a ``typical" state~\cite{Wootters:1990ua,Hayden:2004um,Zyczkowski:2011wo}.  Given a random quantum state, one may then ask how quantum is this state:  does it exhibit classical behavior in the large-number-of-particles regime? If one had access to a black box that prepared random states, how much would it be worth and what would its applications be? These questions and more motivate the present study.

When considering the set of pure states living in an $N$-dimensional complex vector space, the manifold of physical states is the projective space $\mathbb{C}\mathbf{P}^{N}$~\cite{Bengtsson:2017aa}, wherein there is a unique normalized measure invariant under all unitary transformations: the associated Haar measure~\cite{Alfsen:1963wb}.  One can reasonably call this measure the uniform distribution over the unit sphere~\cite{Zyczkowski:1994ua}. 

The geometrical properties of quantum states are essential to understanding where classical intuitions break down and where the next quantum advantage may lie~\cite{Mielnik:1968tp,Braunstein:1994ug,Nielsen:2006vt}.  In phase space, localization is a clear signature  of the state quantumness~\cite{Goldberg:2020tq,Rudzinski:2024aa}, as it has been demonstrated with several indicators, such as R\'enyi-Wehrl entropies~\cite{Gnutzmann:2001wk}, inverse participation ratio~\cite{Wobst:2003vx}, or Husimi extrema~\cite{Marian:2020aa}. However, while these notions provide effective analytical tools, the geometry of $\mathbb{C}\mathbf{P}^{N}$ is not very intuitive and so it is difficult to visualize and to gain insights into the nature of those states.

Here, we propose a novel class of random states based upon a mapping onto a many-qubit system (which, in some cases, might be fictitious).  Actually, any system with a finite-dimensional Hilbert space of dimension $N= 2S+1$ can be thought of as a spin $S$~\cite{Ganczarek:2012un}. The space manifold $\mathbb{C}\mathbf{P}^{N}$ admits an appealing representation due to Majorana~\cite{Majorana:1932ul}, which maps any pure state onto $2S$ points on the unit sphere $\mathcal{S}_{2}$; the state's stellar representation. This picture makes it natural to consider the states associated with random points on the sphere. We conduct a detailed study of these states from an SU(2)-invariant perspective. Notably, we find that, in the limit of a large number of particles, they do not converge to a classical state and preserve their quantumness.

In addition, these new random Majorana (RM) states belong to the symmetric (bosonic) subspace. It has been recognized that symmetric states offer significant advantages from a metrological perspective~\cite{Hyllus:2010aa,Benatti:2013aa,Apellaniz:2015aa,Oszmaniec:2016wo}.  We find numerical evidence that the usefulness of RM states for rotation-sensing-style tasks, like magnetometry and ellipsometry, is robust against the loss of a finite number of particles and dephasing. This is in stark contrast to other relevant states, such as Greenberger-Horne-Zeilinger (GHZ) states~\cite{Greenberger:1989aa} (equivalent to the NOON states~\cite{Dowling:2008aa} in optical interferometry),  which completely lose their (otherwise ideal)  sensitivity upon loss of just a single particle. Finally, we experimentally generate these states by utilizing light's orbital angular momentum and implement arbitrary unitaries using a multiplane light conversion setup to showcase rotation sensing.

This paper is organized as follows. In Sec.~\ref{sec:Majcon} we address the basic notions of the Majorana stellar representation and its physical interpretation. In Sec.~\ref{sec:Rancon}  we raise the ideas behind RM constellations and characterize their properties by resorting to a multipole expansion. Subsequently, in Sec.~\ref{sec:comp} we use these multipoles to compare our method with alternative ways of obtaining random states. In Sec.~\ref{sec:metrology}  we introduce the setting of quantum parameter estimation. Utilizing the quantum Fisher information, we investigate the performance and robustness of the RM states under the influence of noise and particle loss. Finally, in Sec.~\ref{sec:expt}  we introduce the experimental setup used to study these states with respect to rotation sensing and analyze the obtained  results. We conclude our work in Sec.~\ref{sec:conc}.

\section{Majorana constellations}
\label{sec:Majcon}

Let us consider a pure spin-$S$ state $\ket{\psi}$ living in the $2S+1$-dimensional  Hilbert space $\mathcal{H}_{S}$ spanned by the standard angular momentum basis $\{ |S, m\rangle \mid m= -S, \ldots, S\}$, which is the carrier of the irreducible representation (irrep) of spin $S$ of SU(2). This space is isomorphic to $\mathbb{C}^{2S+1}$, but  since any two vectors in $\mathcal{H}_{S}$ differing by a phase represent the same physical state,  the manifold of physical states is the projective space $\mathbb{C}\mathbf{P}^{2S}$~\cite{Bengtsson:2017aa}. 

The merit of Majorana was to show that points in $\mathbb{C}\mathbf{P}^{2S}$ are in one-to-one correspondence with unordered sets of (possibly coincident) $2S$ points on the unit sphere $\mathcal{S}_{2}$. In other words,  spin-$S$ states can be obtained as fully symmetrized states of a system of $2S$ spins $1/2$ (or qubits). These systems have many physical applications, ranging from quantum computation to quantum sensing and metrology~\cite{Bouchard:2017aa,Chryssomalakos:2017tg,Goldberg:2018uq,Martin:2020vu,Goldberg:2021uw}. This idea is also at the heart of the Schwinger map~\cite{Schwinger:1965kx,Chaturvedi:2006vn}, which realizes the set of angular momentum operators in terms of polynomials of bosonic operators. 

There are various ways to see why this is so, but probably the most direct one is to notice that the state $\ket{\psi}$ can always be written as
\begin{equation}
    \ket{\psi}= \frac{1}{\mathcal{N}} \prod_{i=1}^{2S}  a_{- \mathbf{u}_{i}}^\dagger \ket{\mathrm{vac}} ,
    \label{eq:psi with Majorana}
\end{equation} 
where $\mathbf{u}_{i}$ is a unit direction of spherical coordinates $(\theta_i,\phi_i)$, the rotated bosonic operators are 
\begin{equation}
a_{\mathbf{u}}^{\dagger} = \cos(\theta/2) \, a_{+}^{\dagger} +  {e^{i \phi}} \, \sin (\theta/2)  \, a_{-}^{\dagger} \, ,
\end{equation}
with  $a_{+}^{\dagger}$ and $a_{-}^{\dagger}$ creating excitations in two modes (denoted by $+$ and $-$) from the two-mode vacuum $\ket{\mathrm{vac}}$, and $\mathcal{N}$ is a normalization factor of no interest for our purposes here and whose explicit expression can be found, e.g., in Ref.~\cite{Liu:2014ut}. 

The set of $2S$ (non-necessarily distinct) unit vectors $\{ \mathbf{u}_{1}, \ldots, \mathbf{u}_{2S}\}$ defines the Majorana constellation of the state. Alternatively,  the state~\eqref{eq:psi with Majorana} can be expressed as 
\begin{equation}
\ket{\psi} = P_{\psi} (a_{+}^{\dagger}, a_{-}^{\dagger}) \ket{\mathrm{vac}} \, ,
\end{equation} 
where $P_{\psi} (a_{+}^{\dagger}, a_{-}^{\dagger})$ is a homogeneous polynomial of degree $2S$ in the variables $a_{+}^{\dagger}$ and $a_{-}^{\dagger}$ that can be factorized (up to an unessential factor) in the above form.

In particular, the states 
\begin{equation}
\ket{\mathbf{n}} = \frac{1}{\sqrt{(2S)!}} (a_{\mathbf{n}}^\dagger)^{2S} \ket{\mathrm{vac}}
\end{equation} 
are precisely the spin-$S$ coherent states (CS)~\cite{Perelomov:1986ly}. The definition shows that the associated constellations consist of only one single point in the antipodal direction $- \mathbf{n}$. It is natural to introduce the CS representation by $\psi (\mathbf{n}) \equiv \braket{\mathbf{n}|\psi}$: this is a {\emph{bona fide} wave function over the unit sphere $\mathcal{S}_{2}$.} Simple algebraic manipulations yield
\begin{equation}
\psi (\mathbf{n}) = \prod_{i=1}^{2S} \left [ \tfrac{1}{2} (1 - \mathbf{n} \cdot \mathbf{u}_{i}) \; e^{i \Sigma (\mathbf{n}, - \mathbf{u}_{i})} \right ] \, , 
\end{equation}
where $\Sigma (\mathbf{n}, - \mathbf{u}_{i})$ is the oriented area of the spherical triangle with vertices $(\mathbf{z}, \mathbf{n}, - \mathbf{u}_{i})$,  with $\mathbf{z}$ the unit vector in the direction of the axis $Z$. This confirms that the Majorana constellation consists of the zeros of $\psi (\mathbf{n})$.  

The CS wavefunction induces a probability distribution 
\begin{equation} 
Q_{\psi} (\mathbf{n}) \equiv \lvert \braket{\mathbf{n}|\psi} \rvert^{2} \, ,
\end{equation} 
which is the Husimi function~\cite{Husimi:1940aa}, and whose zeros are also the Majorana constellation. 

Because the Majorana representation facilitates a useful geometrical interpretation of quantum states, it has found an increasing number of applications in recent years, with prominent examples being polarimetry and magnetometry~\cite{Goldberg:2019vj}, Bose-Einstein condensates~\cite{Lian:2012ux,Cui:2013wp}, Berry phases~\cite{Hannay:1998vg,Hannay:1998vb,Bruno:2012aa}, and studies of entanglement~\cite{Liu:2016uy}.

\section{Random Majorana constellations}
\label{sec:Rancon}

Since points on the sphere $\mathcal{S}_{2}$ correspond uniquely to pure states via the Majorana representation, one is immediately led to the simplest idea of how to generate RM states: they correspond to sets of random points on $\mathcal{S}_{2}$. Interestingly, the question of distributing points uniformly over a sphere has inspired substantial mathematical research~\cite{Saff:1997aa,Brauchart:2015aa,Alishahi:2015ug}, as has the question of random polynomials~\cite{Bharucha-Reid:1986vr,Hannay:1996wl,Prosen:1996vi,Bogomolny:1996tw}, in addition to attracting the attention of physicists working in a variety of fields.

We randomize each of the spherical coordinates $(\theta_i,\phi_i)$ independently using the Haar measure $\sin\theta_i\, d\theta_i\, d\phi_i/4\pi$ for $\mathcal{S}_{2}$. Equivalently, the resulting RM states can be viewed as arising from the action of a random operator $U\in\mathrm{SU}(2)^{\otimes 2S}$ on the ground state of $2S$  qubits, followed by projection onto the symmetric subspace and normalization. This is fundamentally tied to the most robust deterministic technique for creating arbitrary bipartite states of light, which uses beam splitters and post selection to sequentially add a photon's  coordinates  on its Poincar\'{e} sphere as a new point to a state's existing Majorana constellation~\cite{Fiurasek:2002wp,Kok:2002ta}. When the states of the single photons are randomized, the resulting state immediately takes the form of Eq.~\eqref{eq:psi with Majorana} with random spherical coordinates. This basic scheme can then be used to create random states in the light's polarization degree of freedom for tasks like polarimetry.

\begin{figure}[t]
  \centerline{\includegraphics[width=0.95\columnwidth]{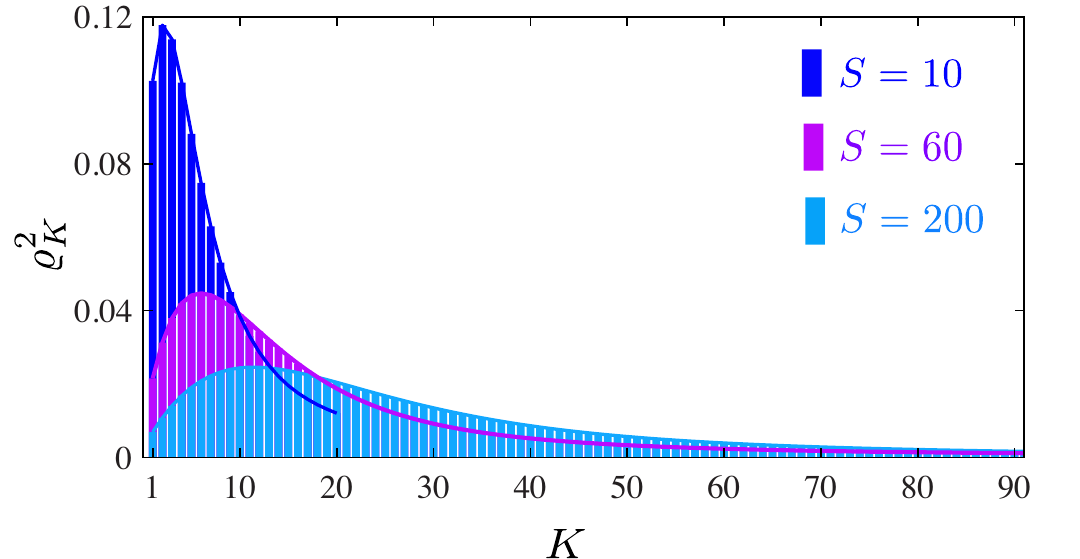}}
  \caption{Mutipole distribution $\varrho_{K}^{2}$ for RM states with the values of $S$ indicated in the inset.}
  \label{fig:RanvarS}
\end{figure}

The intrinsic SU(2) symmetry suggests using the germane notion of multipoles~\cite{Fano:1959ly,Blum:1981rb} to characterize the resulting states. To this end, we expand the density matrix of the system as 
\begin{equation}
\varrho = \sum_{K=0}^{2S}\sum_{q=-K}^K\varrho_{Kq} T_{Kq} \, ,
\end{equation} 
where $T_{Kq}$ are the spherical tensor operators, defined as~\cite{Blum:1981rb}
\begin{equation}
T_{Kq}=\sqrt{\frac{2K+1}{2S+1}}\sum_{m,m^\prime =-S}^S C_{Sm,Kq}^{Sm^\prime}|S,m^\prime\rangle\langle S,m| \, ,
\end{equation} with $C_{Sm, Kq}^{Sm^{\prime}}$ being the Clebsch-Gordan coefficients that couple a spin $S$ and a spin $K$ \mbox{($0 \le K \le 2S$)} to a total spin $S$.  These tensors are an orthonormal basis
\begin{equation} 
\Tr(T_{Kq}\, T_{K'q'}^\dagger)=\delta_{KK'}\delta_{qq'}
\end{equation}
that transform appropriately under rotations $R(\bm{\Omega}) \in$ SU(2):
\begin{equation}
R (\bm{\Omega}) \, T_{Kq} \, R^\dagger (\bm{\Omega}) = \sum_{p} D_{pq}^{S}(\bm{\Omega}) \, T_{Kp} \, ,
\end{equation}
where $D_{mn}^{S}(\bm{\Omega})=\langle S,m|R(\bm{\Omega})|S,n\rangle$ are the usual Wigner $D$ matrices~\cite{Varshalovich:1988xy}.

\begin{figure}[t]
  \centerline{\includegraphics[width=0.95\columnwidth]{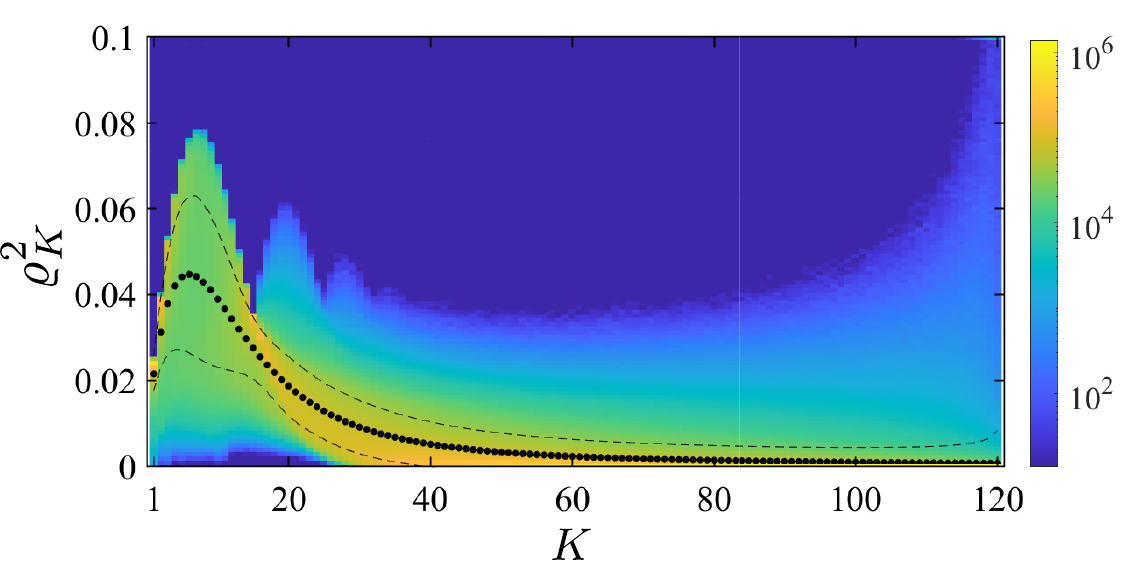}}
  \caption{The dots represent the average multipoles for RM states with $S=60$. The broken lines delimit the associated variances. The density plot in the back (with the logarithmic scale shown at the right) is the number of trials having the corresponding value of the multipole.}
  \label{fig:Var}
\end{figure}

The expansion coefficients are scalar values that are precisely the state multipoles. Alternatively, they can  be computed as~\cite{Bouchard:2017aa}
\begin{equation}
\label{eq:multYkq}
  \varrho_{Kq} = \mathcal{C}_K \int_{\mathcal{S}_2} 
 d^2\mathbf{n} \; Q_{\varrho} (\mathbf{n}) \; {Y}_{Kq}(\mathbf{n}) \, ,
  \end{equation}
where  $\mathcal{C}_K$ is a constant, ${Y}_{Kq}(\mathbf{n})$ the standard spherical harmonics, and  $d^2\mathbf{n} = \sin \theta d\theta d\phi$ the invariant measure on $\mathcal{S}_{2}$. The multipoles thus appear as the standard ones in electrostatics, but replacing the charge density by $Q_{\varrho}(\mathbf{n})$ and distances by directions~\cite{Jackson:1999aa}. They are the $K$th directional moments of the state and, therefore, they resolve progressively finer angular features with increasing $K$. 

Since ${Y}_{Kq}(\mathbf{n})$ constitute an orthonormal basis on $\mathcal{S}_{2}$, we can immediately invert Eq.~\eqref{eq:multYkq}:
\begin{equation} 
Q_\varrho(\mathbf{n}) = \mathcal{C}_{K}^{-1} \sum_{Kq} \varrho_{Kq} Y_{Kq}(\mathbf{n}) \, , 
\end{equation}
so the multipoles represent the state's $Q$ distribution on the sphere in its natural basis of spherical harmonics.  As the $Q$ function determines the Majorana constellation, these coefficients $\varrho_{Kq}$ can also be directly computed from the  constellation, as was recently explained in Ref.~\cite{Romero:2024aa}.

For each value of $S$, we average over $1.5 \times 10^{6}$ samples.  In Fig.~\ref{fig:RanvarS} we show the resulting multipoles  plotted in terms of the multipole squared length 
\begin{equation}
\varrho_{K}^{2} = \sum_{q=-K}^{K} | \varrho_{Kq} |^{2} \,.
\end{equation} 
which are SU(2) invariant and thus serve as an effective tool for studying the states, treating all states related by an SU(2) transformation (a rigid rotation of the constellation) as equivalent. As we can see, only multipoles with small $K$ contribute significantly. The maximally contributing multipole $K_{\mathrm{max}}$ smoothly increases with $S$: a least-squares fitting gives that, for large $S$, $K_{\mathrm{max}} = a \sqrt{S}$, with $a \simeq 0.8$. 

To gain further insight into this behaviour, in Fig.~\ref{fig:Var} we plot the average multipoles for RM states with $S=60$. The broken lines delimit the corresponding variances, which are clearly not uniform. Since the individual multipole distributions are non-Gaussian, the variances give only partial information. To complete the picture, in the background we display a density plot representing the number of states with a given value of the $K$th multipole. Notice that we use a logarithmic scale in order to better appreciate the behaviour of higher $K$s, which have exceedingly small values. We see the emergence of a striking multipeaked structure. The multipoles with significant yellow areas (i.e., a strong concentration of samples) are those with less variance.

\begin{figure}[t]
  \centerline{\includegraphics[width=0.90\columnwidth]{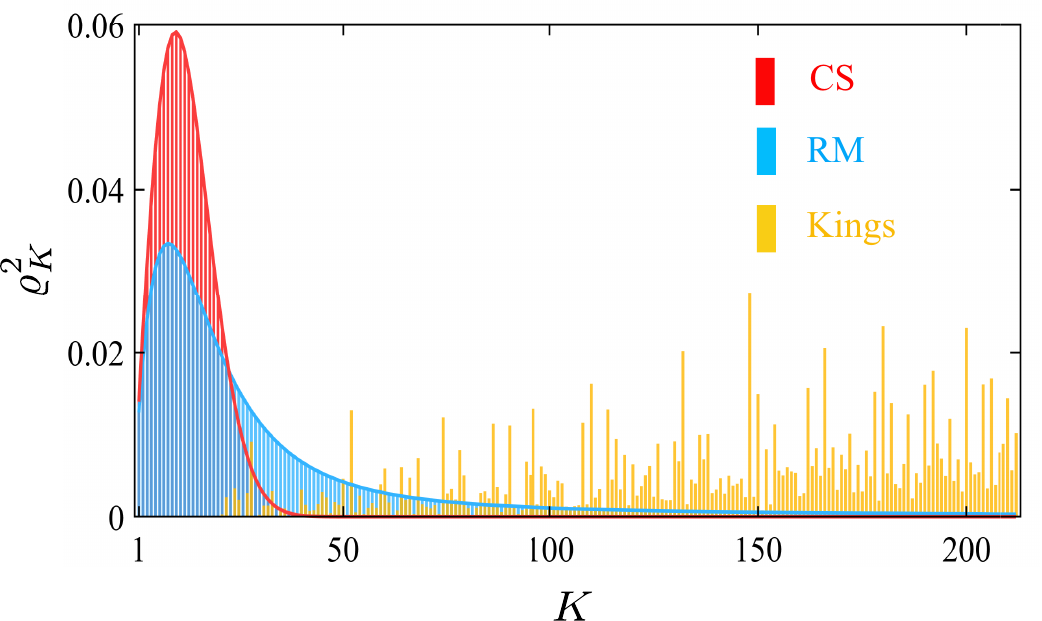}}
  \caption{Multipole distribution for CS, RM, and Kings of Quantumness for $S=106$. The colors in the inset indicate the corresponding states.}
  \label{fig:Multvar}
\end{figure}

One sensible set of quantities in this scenario is the set of cumulative multipole distributions, defined as 
\begin{equation}
A_{M}= \sum_{K=1}^{M} \varrho_{K}^{2} \, .
\end{equation} 
Note that we ignore the 0th multipole as it corresponds only to normalization. For CS this quantity reaches the value
\begin{equation}
  \label{eq:Aksu2}
  {A}_{M}^{\mathrm{CS}} = \frac{2S}{2S +1} -
  \frac{[\Gamma (2S + 1)]^{2}}{\Gamma (2S-M) \Gamma (2S + M +2)} \, ,
\end{equation}
and it has been proven that this is indeed maximal for every $M \in \{ 1,\ldots,2S \}$~\cite{Bjork:2015vg}. At the opposite extreme we have states whose multipoles vanish up to the highest $M$: they have been dubbed as Kings of Quantumness~\cite{Bjork:2015vv} and they are maximally unpolarized. For each total spin $S$, there exists a maximal order $M$ to which a state can be unpolarized; the state(s) satisfying this condition of $A_M=0$ are the Kings. Therefore,  $A_M$ is a good measure of the quantum properties of a state through its \emph{hidden polarization} features~\cite{Klyshko:1992tn} that are stored in high-order multipoles.

\begin{figure}[t]
  \centerline{\includegraphics[width=0.90\columnwidth]{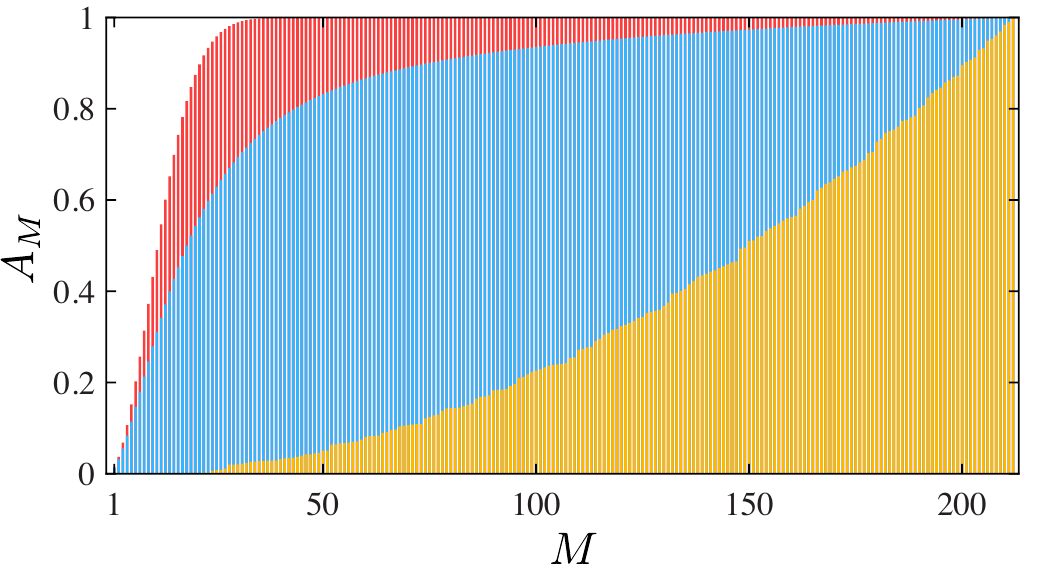}}
  \caption{Cumulative multipole distribution for CS, RM, and Kings of Quantumness states for $S=106$. The colors are the same as in Fig.~\ref{fig:Multvar}.}
  \label{fig:Multvarcum}
\end{figure}

To appreciate the different behaviours, in Fig.~\ref{fig:Multvar} we have plotted the multipole distributions for the most quantum (Kings of Quantumness), the least quantum (CS), and RM states for $S=106$. The differences speak for themselves. A RM constellation is markedly different from a classical state: although the maximal contributions arise from roughly the same multipoles for both states (actually, for CS a quick estimate gives $K_{\mathrm{max}} \simeq \sqrt{S}-1/2$), RM states have much heavier tails, thus hiding  their quantum information in higher-order multipoles than CS. This makes them useful for metrological applications, even in the large-$S$ limit, where quantum effects may be expected to vanish. Additionally, such RM states are far from the most quantum states, certifying the rarity of the Kings of Quantumness and the effort required for creating them. This behaviour is confirmed by the cumulative multipole distribution $A_{M}$ for the same states, as plotted in Fig.~\ref{fig:Multvarcum}.

\section{Comparison to other random distributions}
\label{sec:comp}

If a pure state is expressed in the angular momentum basis as $|\psi\rangle=\sum_m \psi_m |S,m\rangle$, one could instead consider states with randomized probability amplitudes $\psi_{m}$. This can be achieved with a random unitary $U \in\mathrm{SU}(2S+1)$ acting on the state, which is often called the circular unitary ensemble (CUE)~\cite{Dyson:1962wh}. These random unitaries were considered in Refs.~\cite{Slomczynski:1998tj} and~\cite{Gnutzmann:2001wk}, who found the R\'enyi-Wehrl entropies of the resulting random CUE states to be highly quantum. Moreover, in Ref.~\cite{Oszmaniec:2016wo} it was demonstrated that such random CUE states are useful and robust for metrology. 

Actually, we can calculate the expectation value of the cumulative multipole moments $\mathcal{A}_M = \int A_M \,dU$ for a variety of normalized Haar measures $dU$, through
\begin{equation}
\mathcal{A}_M  =\sum_{K=1}^M\sum_{q=-K}^K\frac{2K+1}{2S+1}\sum_{m, m^{\prime}=-S}^S C_{Sm,Kq}^{Sm+q} C_{Sm^\prime,Kq}^{Sm^\prime+q}I_{m,m^\prime,q},
\end{equation}
with $ I_{m,m^\prime,q}=\int \psi_{m+q}\psi_m^\ast \psi_{m^\prime+q}^\ast \psi_{m^\prime} dU$, and $C_{S_{1}m_{1}, S_{2} m_{2}}^{Sm}$ the Clebsch-Gordan coefficients~\cite{Varshalovich:1988xy} that  vanish unless the usual angular momentum coupling rules are satisfied: $m_1+m_2=m$, $ 0 \leq K \leq 2S$, and $ -K\leq q \leq K$.

For example, if we express our random unitaries by $U\in\text{SU}(2S+1)$,  these integrals can be obtained exactly using random-matrix theory~\cite{Weingarten:1978uy,Gorin:2008ux} (see Supplemental Material). But there is another intuitive method: all of the nonzero integrals take the form $\int |\psi_m|^2|\psi_n|^2 dU$ and the distribution for each $\psi_i$ is the same, so we know immediately that $I_{m,m^\prime,q}\propto \delta_{q,0}+\delta_{m,m^\prime}$, such that 
\begin{equation}
    \mathcal{A}_{M,\; \mathrm{SU}(2S+1)} =\frac{M(M+2)}{(2S+1)(2S+2)}.
    \label{eq:AM average qudit}
\end{equation} 
This means that states with random coefficients $\psi_m$ have $\sum_{q=-K}^K|\varrho_{Kq}|^2\propto 2K+1$, making them much more quantum than RM states  and according with other results for this distribution~\cite{Slomczynski:1998tj,Gnutzmann:2001wk}. This is in stark contrast to the notion that both forms of randomness approach each other in the limit of large $S$ in terms of the distance between arbitrary pairs of random states~\cite{Zyczkowski:2001td}, stressing the differences maintained between the forms of randomness for all but $S=1/2$.

\begin{figure}[t]
  \centerline{\includegraphics[width=0.90\columnwidth]{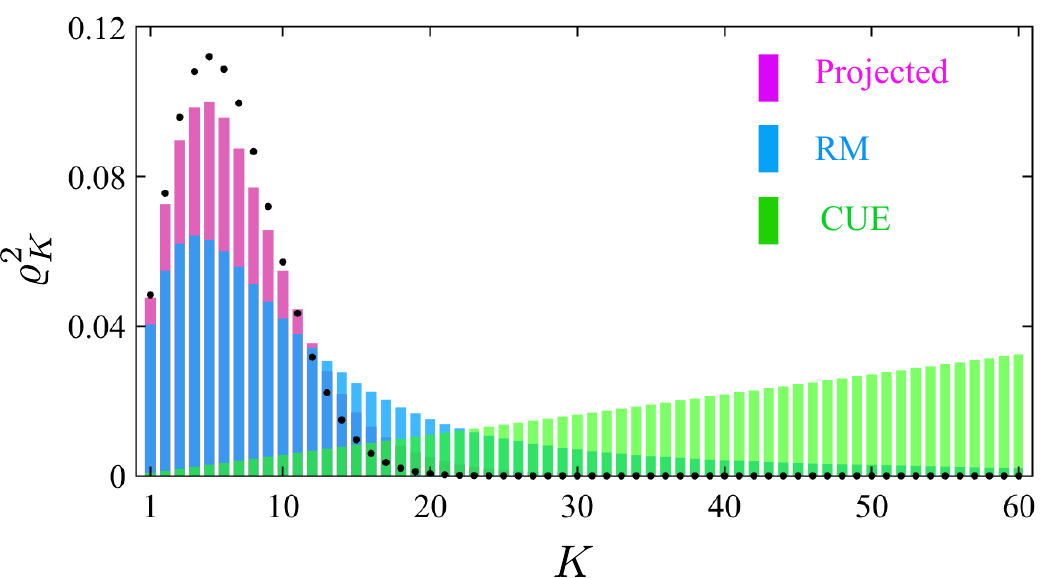}}
  \caption{Multipole distribution for RM, CUE, and projected random qubits, all of them for  $S=30$. For comparison, the dots represent the multipoles for a CS.}
  \label{fig:diff}
\end{figure}

Another possibility of obtaining a random spin-$S$ state is to take a state of $2S$ random qubits. We employ the same procedure as used for the RM, beginning with the $2S$ qubits, but without normalizing the states initially. The resulting state is normalized at the end, ensuring that each state carries a weight proportional to its overlap with the symmetric subspace. This contrasts with the RM, where all states are assigned equal weight. This is not a unitary operation, instead creating the states from Eq.~\eqref{eq:psi with Majorana} with the replacement of the normalization constant by a state-independent factor $\mathcal{N}\to\sqrt{\left(2S\right)!}$, but it facilitates an analytical calculation whose cumbersome expression we show in the Supplemental Material. 

Because each state in the ensemble has a different normalization due to having inhomogeneous likelihoods of being produced, the statistical properties derived for them should be understood as being weighted by the probabilities of finding the different projected states. These probabilities are proportional to $\mathcal{N}$ and their normalization can be found from the 0th order multipole. After normalizing the final results, we find that their multipolar distribution is very similar to that of CS. Intuitively, one can conclude that it is far more likely for such a projection method to produce a state close to CS than to produce a highly quantum state. The RM states may be thought of as states randomly chosen after the projection method succeeds, while the projected states are the result of randomization prior to the projection. In other words, if the RM can be thought of as a die with $n$  equally likely sides, the projected states are lthe same die, but with each side having a different probability.

\begin{figure}[t]
  \centerline{\includegraphics[width=0.88\columnwidth]{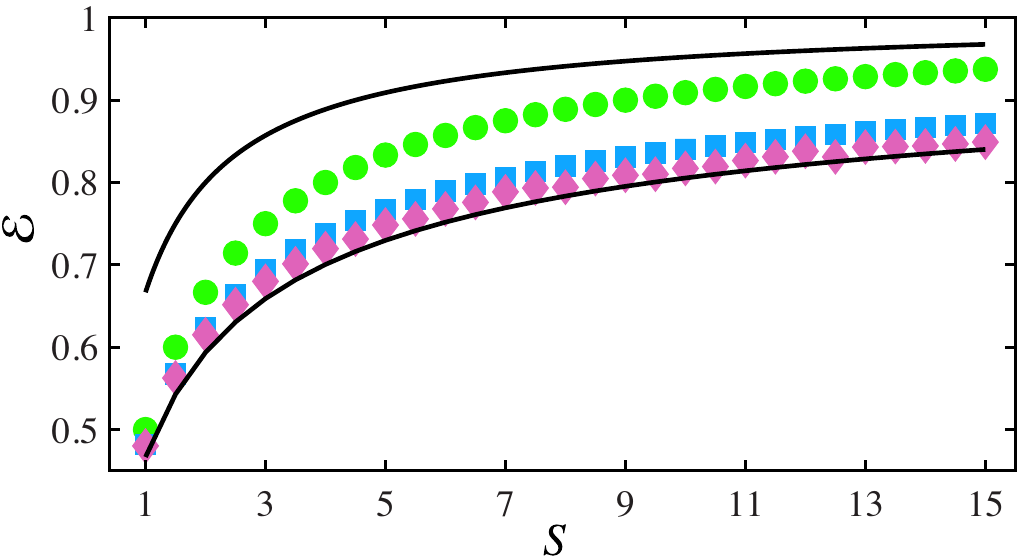}}
  \caption{Quantumness $\mathcal{E}$ for the same random states considered in Fig.~\ref{fig:diff} as a function of $S$. The two continuous lines give the limit values of $\mathcal{E}$: the upper one is $2S/(2S+1)$, whereas the lower one corresponds to the CS.}
  \label{fig:Enta}
\end{figure}

In Fig.~\ref{fig:diff} we display the conspicuous differences between the multipole distributions for these different randomized states.

The conclusion from the discussion thus far is that classical states convey their information in lower multipoles, whereas the opposite occurs for extremal quantum states. However, to assess this behaviour in a quantitative way, we need a proper measure of quantumness. Since our analysis has been largely based on multipoles, we will use a recent proposal defined precisely in terms of them~\cite{Goldberg:2021tc}; viz.,
\begin{equation}
\mathcal{E}(\varrho) = 1 - 
\sum_{K=0}^{2S} \frac{\varrho_{K}^2}{2K+1} \, .
\label{eq:ent}
\end{equation} 
$\mathcal{E}(\rho)$ takes an entanglement monotone for a particular decomposition of Hilbert space, the linear entropy of the pure state when decomposed over the modes $a_{\mathbf{u}}$ and $a_{-\mathbf{u}}$, then averages this over all possible two-mode decompositions that come from rotating the unit vector ${\mathbf{u}}$. We stress though that this  metric  is an entanglement monotone only for pure states. Equation~\eqref{eq:ent} provides  a mode-decomposition-agnostic measure of quantumness that is entanglement averaged over easy-to-perform entangling operations. Larger values of $\mathcal{E}$ signify more quantum states. This quantity emerges when looking at symmetric superpositions of two-mode states: it turns out that they can be entangled or separable, but this property can change after the Majorana constellation undergoes a rigid rotation. To properly account for this possibility, one can consider their linear entropy of entanglement averaged over all rotated partitions of the two-mode Hilbert space: the final result is precisely~\eqref{eq:ent}.

In Fig.~\ref{fig:Enta} we calculate this measure for the different states discussed before for various values of $S$. We see that RM constellations are more quantum than states of random qubits projected onto the symmetric subspace, but less quantum than CUE states  with random coefficients in the angular momentum basis. Impressively, the CUE states have $\mathcal{E}_{\mathrm{CUE}}=2S/(2S+2)$ on average, which is very close to the maximum value for a single pure state: $2S/(2S+1)$. The lower continuous line in the figure corresponds to the values of $\mathcal{E}$ for CS, which are the least quantum ones.

This confirms the fact that the overwhelming majority of CUE random states are extremely close to the maximally entangled state~\cite{Hayden:2006aa} and that a significant portion of the RM states are entangled, which seems very counterintuitive. Moreover, we observe that for many functions defined over $\mathcal{H}_{S}$, the  majority of vectors take a value of the function very close to the average value as $S \rightarrow \infty$. This observation, collectively, is referred to as the concentration of measure phenomenon~\cite{Singh:2016aa}.

\section{Metrology with random states}
\label{sec:metrology}

We look next at the possibility of sensing rotations using random states. A general rotation is characterized by three parameters: the two angular coordinates fixing the rotation axis $\mathbf{u}(\Theta,\Phi)$ and the angle  $\omega$ rotated around that axis~\cite{Grafarend:2011aa}. In the space $\mathcal{H}_{S}$ the action of this rotation is represented by the operator~\cite{Cornwell:1984aa} 
\begin{equation}
R ( \bm{\Omega}) = \exp( i \omega \, \mathbf{S} \cdot \mathbf{u}) \, , 
\end{equation} 
where we have used the notation $\bm{\Omega} (\omega,\mathbf{u}) = (\omega,\Theta, \Phi)$ to denote the three parameters and $\mathbf{S}$ is the vector comprising the three components of the angular momentum; the generators of the algebra $\mathfrak{su}(2)$. 

The restriction of working in a single irrep is reasonable, since maximal precision will be obtained by concentrating all of the resources into a single subspace corresponding to the average total number of particles. This is true in the local regime, where prior information about the parameter in question is known~\cite{Kolenderski:2008aa}, in sharp contrast to the global regime, where minimum error in estimating a completely unknown parameter requires coherences between irreps~\cite{Chiribella:2004aa}.

For the time being, we assume the rotation axis $\mathbf{u}$ to be known; the task is thus to estimate the rotation angle $\omega$.  A canonical scenario requires  $\omega$ to be imprinted on a (preferably pure) probe state $\ket{\psi}$, in which the latter is shifted by applying a rotation $R (\omega) \in \mathrm{SU}(2)$ that encodes the angle $\omega$. A set of measurements is then performed on the output state $\ket{\psi_{\omega}} = {R} (\omega ) \ket{\psi}$, with the measurements represented by a positive operator-valued measure (POVM)~\cite{Helstrom:1976aa} $\{ \Pi_{x}\}$, where the POVM elements are labeled by an index $x$ (discrete or continuous) that represents the possible outcomes of the measurement according to Born's rule $p(x | \omega) = \bra{\psi_{\omega}} \Pi_{x} \ket{\psi_{\omega}}$. Afterward, what remains is to infer the angle  via an estimator $\hat{\omega}$~\cite{Kay:1993aa}, whose  performance is usually assessed in terms of the variance. The ultimate limit for any possible POVM is given by the quantum Cram\'er-Rao bound (QCRB), which reads~\cite{Braunstein:1994aa}
\begin{equation}
\Var_\psi (\hat{\omega})  \geq \frac{1}{\nu  \; \mathsf{Q}_{\psi}( \omega )} \, ,
\end{equation} 
where $\nu$ is the number of independent times the experiment is repeated. To assess the ultimate sensitivity per experimental trial, we take henceforth $\nu = 1$. Here, $\mathsf{Q}_{\psi}( \omega )$ is the quantum Fisher information (QFI), which  depends exclusively on the initial probe state. We briefly recall that an  explicit way to compute $\mathsf{Q}_{\varrho} ( \omega )$ is as $\mathsf{Q}_{\varrho} ( \omega ) = \Tr (\varrho_{\omega}  L^{2}_{\omega})$ and $L_{\omega}$ is the so-called symmetric logarithmic derivative, defined implicitly via~\cite{Szczykulska:2016aa,Sidhu:2020aa,Albarelli:2020aa}
\begin{equation}
\frac{\partial {\varrho}_{\omega}}{\partial \theta} = 
\case{1}{2} \{ \varrho_{\omega}, L_{\omega} \} \, , 
\end{equation}
and $\{ \cdot, \cdot \}$ stands for the anticommutator $\{ A, B \} = A B + B A$. For pure initial states we have the simple result
\begin{equation}
\mathsf{Q}_{\psi} ( \omega ) = 4 \, \Var_{\psi} (\mathbf{S}\cdot \mathbf{u} ) \, .
\end{equation} 
although generalizations to general density matrices are possible~\cite{Fiderer:2019aa}. Convexity of the variance is responsible for a single irrep conferring maximal precision per number of particles. The QFI is thus maximal for pure states that maximize the variance of the generator $\mathbf{S}\cdot \mathbf{u}$: these are the states $ \ket{\mathbf{u}} + \ket{- \mathbf{u}})/\sqrt{2}$, which are the rotated versions of the time-honoured NOON states~\cite{Dowling:2008aa}, defined as  
\begin{equation} 
{\ket{\mathrm{NOON}} = \frac{1}{\sqrt{2}} (\ket{SS} + \ket{S,- S}) \, .}
\end{equation}

One may be interested in a probe state with the best average performance for any $\mathbf{u}$ rather than creating a distinct state for every axis. {To this end, we consider the average QFI defined as}
\begin{equation}
    \bar{\mathsf{Q}}_{\varrho} (\omega)  = \frac{1}{4\pi} 
    \int_0^\pi {\! \! d\Theta}\, \sin\Theta \, \int_0^{2\pi} \!\! d\Phi\,\mathsf{Q}_{\varrho} (\omega)  \, .
\end{equation} 
This average has been used in different issues of quantum information~\cite{Toth:2014aa}. An exemplary context is magnetometry, where the magnitude of the magnetic field being sensed can be more important than its orientation. Then, $\bar{\mathsf{Q}}_{\varrho} (\omega)$  gives an upper bound on the attainable precision for a quantum state $\varrho$, if the direction of the magnetic field is chosen randomly based on a uniform distribution.

\begin{figure}[t]
  \centerline{\includegraphics[width=1.05\columnwidth]{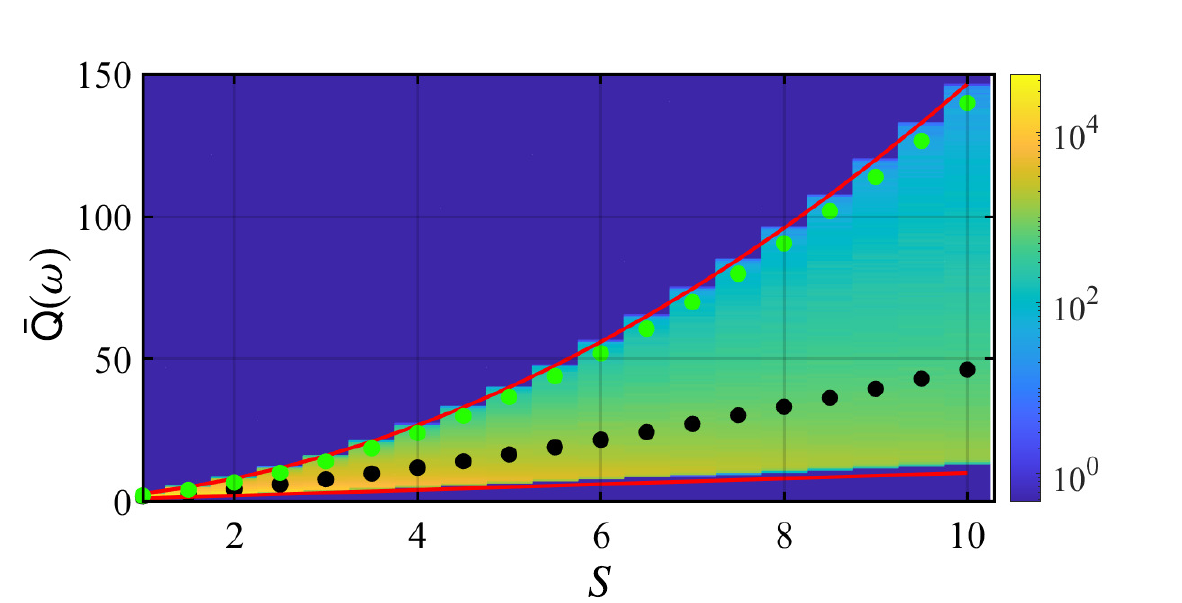}}
  \caption{Average QFI as a function of the spin $S$. The black (green) dots denote the average value (over 60k trials) of the RM (CUE) states. The density plot in the back is the number of trials having the corresponding value of the average QFI for RM.  The red lines indicate the maximum and minimum achiveable values of the QFI, corresponding to first-order unpolarized and CS, respectively.}
  \label{fig:QFIave}
\end{figure}

The maximum average QFI is achieved by states whose angular momentum projection vanishes in all directions; i.e., 
\begin{equation}
\Tr (\varrho_{\omega} \mathbf{S} ) =\mathbf{0} \,, 
\end{equation} 
with a maximum value $\bar{\mathsf{Q}}_{\mathrm{max}} =\case{4}{3}S(S+1)$. They correspond to first-order unpolarized states (which, among others, include the NOON states). In this sense, the most sensitive probe states are those whose classical angular momentum features are hidden~\cite{Goldberg:2020ac,Serrano-Ensastiga:2024aa}.

\begin{figure}[t]
  \centerline{\includegraphics[width=\columnwidth]{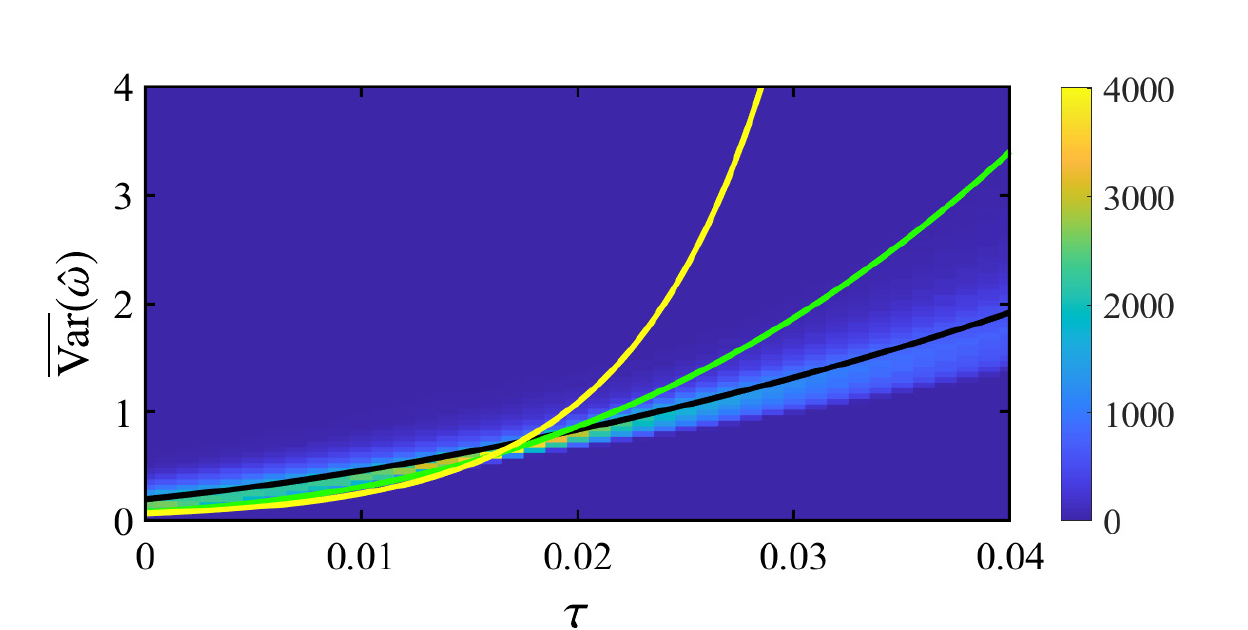}}
  \caption{Average variance for RM (black line), CUE (green line) and King of Quantumness (yellow line) states as a function of the dimensionless time $\tau = \gamma t$
   when the system experiences dephasing according to the model in Eq.~\eqref{eq:deph}
   . In all cases, $S=3$. The density plot in the back is the same as in previous plots.}
  \label{fig:QFIdeph}
\end{figure}

In Fig.~\ref{fig:QFIave} we plot the average $\bar{\mathsf{Q}}_{\varrho} (\omega)$ for both, RM and CUE  states as a function of $S$. We also include the extremal values of this quantity, which are for first-order unpolarized and CS, respectively. The random states always lie in between these two lines, although CUE are closer to optimal. Some of the random states have QFI close to the maximum values, but most of them  are concentrated below its average value.

In view of these results, one might wonder about the usefulness of random states. Intuition suggests that they should be more robust against imperfections than very fragile entangled states, as is the case for CUE states with certain types of loss~\cite{Oszmaniec:2016wo}.  To confirm this point, let us first consider the case of depolarization. We use the model devised in Ref.~\cite{Rivas:2013aa} for an SU(2)-invariant dephasing, which is appropriate for our case. The time evolution of the state is given by
\begin{equation}
\label{eq:deph}
\varrho(t) = \exp(-2 \gamma t \mathbf{S}^2) \sum_{n=0}^{\infty} \sum_{j=x,y,z} \frac{(2 \gamma t)^n}{n!} S_j^n \varrho(0) S_j^n \, ,
\end{equation}
where $\gamma$ is a constant determining the strength of the dephasing, which we set to 1 without loss of generality. The results of the evolution under this dephasing are summarized in Fig.~\ref{fig:QFIdeph} for the case of $S=3$, where we plot the minimum average variance of the angle estimator $\hat{\omega}$ as a function of the dimensionless time evolution (that can be easily converted to, e.g.,  traveled distance), for RM, CUE, and Kings of Quantumness states. The Kings are the optimal states to sense an arbitrary rotation when quantified by the highest quantum Fisher information averaged over all rotations and by the lowest inverse quantum Fisher information when averaged over all rotations~\cite{Goldberg:2021uw}, and, as expected, are very vulnerable to dephasing, and become much worse than the random, not only in their average values, but for most of their individual trials. Interestingly, RM are more robust than CUE and similar behaviour is observed for other dimensions.

\begin{figure}[t]
  \centerline{\includegraphics[width=\columnwidth]{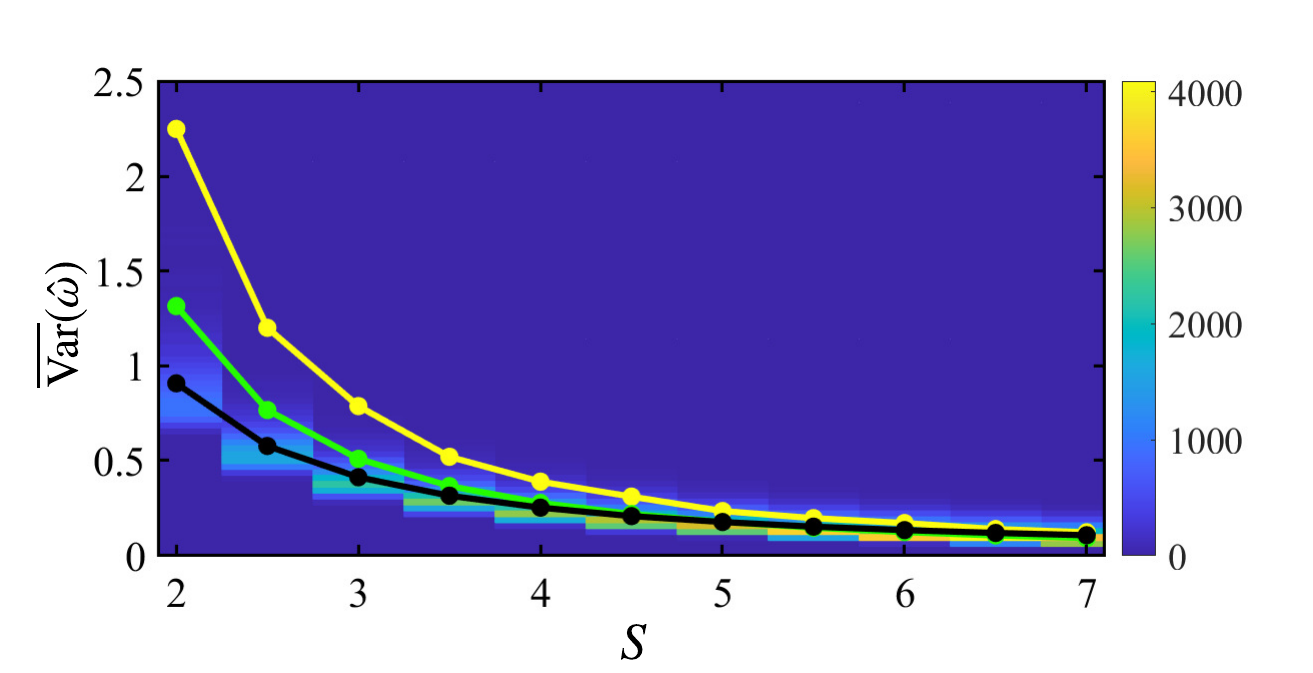}}
  \caption{Average variance for RM (black line), CUE (green line) and King of Quantumness (yellow line) states as a function of $S$ after losing one particle. The density plot in the back is the same as in previous plots.}
  \label{fig:QFIPL}
\end{figure}

We next examine the case where a photon loss occurs. To analyze this scenario, we  employ a simple SU(2)-invariant model that consists of mapping each state to an incoherent superposition of the states resulting from the removal of one particle, which in general is a mixed state in the space of spin $S-\tfrac{1}{2}$. In other words, we create $2S$ states by the removal of each one of the creation operators in  Eq.~\eqref{eq:psi with Majorana} and add them incoherently, to explicitly look at random Majorana constellations that randomly lose a star; this is different from mode-agnostic removal of a particle~\cite{Goldberg:2024aa} and is more in line with a preparation error where one of the inputs is randomly neglected. The results are shown in Fig.~\ref{fig:QFIPL}, for the same states as in the previous Figure. One can observe that for low $S$, random states exhibit again a clear advantage over Kings of Quantumness, albeit it diminishes as $S$ increases.

The measurement saturating the QCRB has been characterized~\cite{Goldberg:2021uw}, but its experimental implementation may be challenging.  Easier is to project the rotated state onto a set of coherent states for various directions and to reconstruct the rotation parameters from these measurements.  Using the same basic principles applied to geographical positioning systems (GPS)~\cite{Hofmann-Wellenhof:2001aa},  five projections are sufficient for this orientation problem, as has been recently demonstrated~\cite{Eriksson:2023aa}. This is particularly useful in the context of random-state metrology, where we seek a single measurement scheme that is useful for all input states; while it cannot generally attain the averaged QFI, due to the QFI allowing for a different measurement for each term being averaged over, it can often perform very well, as demonstrated experimentally below.

\begin{figure*}
    \centering
    \includegraphics[width=1.75\columnwidth]{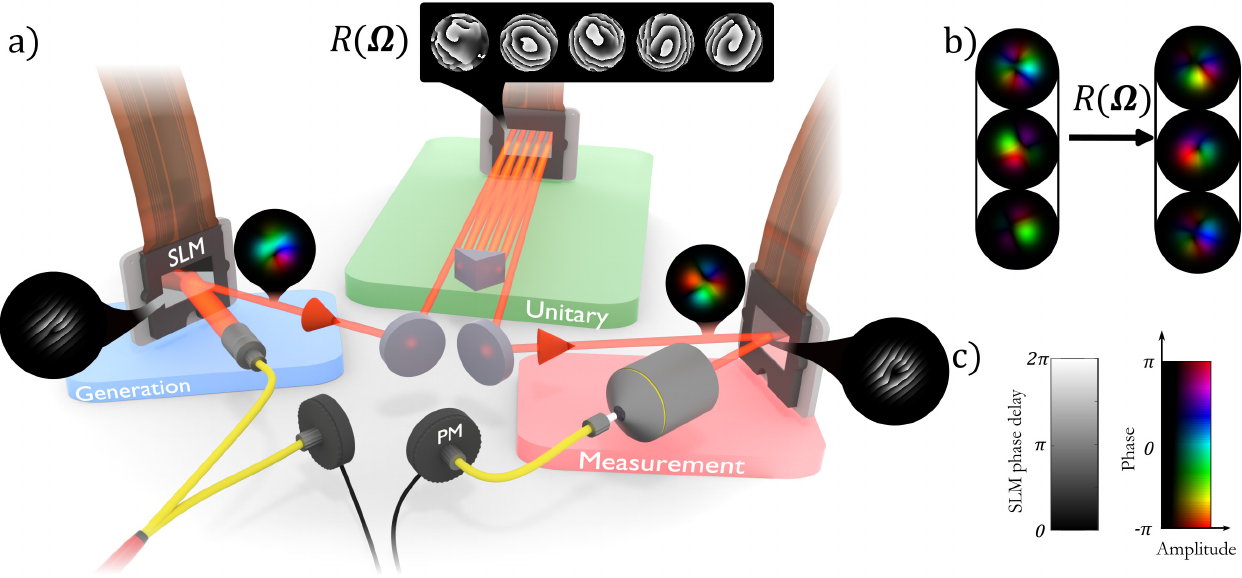}
     \caption{A simplified sketch of the experimental system is shown in a).  The initial laser output is split into two using a fiber beamsplitter with half of its power being directed onto a power meter (PM). The rest is collimated and sent onto the first of three SLMs. The Gaussian beam coming out of the single-mode fiber (SMF, yellow) is shaped into one of the chosen random states by the first SLM. The prepared probe state is then sent to the MPLC device that performs a unitary rotation operation $R ( \bm{\Omega})$.   After the MPLC, the rotated probe state is sent to the third SLM to be measured. The third SLM, together with a final SMF, perform a projective measurement onto a state by first shaping the phase and amplitude of the beam and then coupling the first diffraction order of the SLM into the SMF using a $10\times$ microscope objective~\cite{Bouchard:2018aa}. Non-magnifying $4f$ lens systems imaging each SLM onto the next one are omitted from the figure. The  insets correspond to field structures (in color) and phase mask structures (in gray scale). Colorbars explaining the mode structure and gray-scale colormaps are given in c).  The holographic measurement mask shown in a) filters for one of the chosen coherent states in the 5-dimensional Hilbert space.      The field structure insets in a) show an example of a state with a randomly distributed Majorana constellation for $S=2$ and the same transverse field structure after the rotation unitary. In b), more examples of states with random constellations are shown, along with their rotated counterparts. In the sketch of the experiment, the grating terms have been removed from the phase masks of the unitary and only the field structure at the first-diffraction orders from each phase mask are shown.}
    \label{fig:setup}
\end{figure*}

\section{Experimental results}
\label{sec:expt}

The results of Sec.~\ref{sec:metrology} suggests that random states should perform well in rotation sensing tasks, in systems with imperfections. Hence, as a last step, we set out to test the suitability of random states for rotation sensing in the laboratory, where experimental imperfections can never be avoided. In these measurements, we used multiple different random states in the GPS-like measurement protocol mentioned above.

In our experimental implementation, we used the transverse-spatial degree of freedom of light. More specifically, we construct our Hilbert space $\mathcal{H}_{S}$ from a set of Laguerre-Gauss (LG) modes $\mathrm{LG}_{\ell}^{p}$, only choosing modes with a zero radial index $p = 0$ and orbital angular momentum (OAM) indices $\ell$ corresponding to the $m$ index in the standard angular momentum basis $|S, m\rangle$. This is a standard system for probing SU(2) dynamics, even though it can also be analyzed with classical optics~\cite{DAmbrosio:2013aa}.  The beam radius was around $520~\mu \mathrm{m}$ for the LG modes used to prepare the probe state. For both RM and CUE states, we follow the generation procedure described in Sec.~\ref{sec:Rancon} and \ref{sec:comp}, respectively,  and translate the resulting states to our laboratory encoding using a set of $(2S+1)$ LG modes. We then constructed the rotation $R ( \bm{\Omega})$, within this Hilbert space, using a multiplane light conversion (MPLC) device~\cite{Morizur:2010ua,Labroille:2014aa}. For the measurements, Hilbert spaces with dimensions 5 and 7, corresponding to spins $S=2$ and $S=3$, respectively, were used.

The experimental setup is shown in Fig.~\ref{fig:setup} and can be divided into three distinct parts: probe state generation, unitary rotation of the probe state, and measurement of the rotated state. We used a CW laser at roughly 808.4~nm, for all of the measurements. The random probe states were created as coherent superpositions of the chosen set of LG modes by shaping an initial Gaussian field with an amplitude and phase modulating mask~\cite{Bolduc:2013aa} on a spatial light modulator (SLM, Holoeye Pluto-2). An additional Gaussian correction was added to the masks~\cite{Plachta:2022aa,Hiekkamaki:2022aa}, and all of the SLMs used in the experiment had an added phase profile for aberration correction that was retrieved using the method introduced in Ref.~\cite{Jesacher:2007aa}.

In the second part of the experimental setup, an MPLC system, with five phase modulations, transforms the probe field according to the rotation transformation $R(\bm{\Omega})$. The phase modulations were performed on a single SLM, and each pair of consecutive modulations was separated by $800$~mm of free-space propagation. The MPLC device is capable, in principle, of performing an arbitrary unitary mapping between spatial modes~\cite{Morizur:2010ua, Brandt:2020aa}. 

A set of five phase modulating masks had to be designed, through a process called wavefront matching~\cite{Fontaine:2019aa}, for each of the 10 rotation angles [$(0, 36, \ldots, 324)$ degrees], in both the 5- and 7-dimensional systems. The rotation axis was kept fixed in each dimension. The third and final SLM was used to measure the rotated probe state by projecting it onto a transverse field structure~\cite{Bouchard:2018aa}.

\begin{figure*}
    \centering
    \includegraphics[width=2\columnwidth]{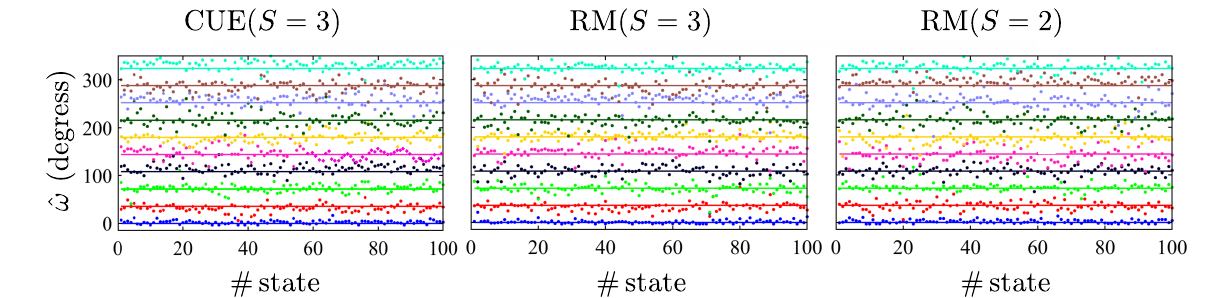}
    \caption{The dots are the estimated values of the rotation angle $\hat{\omega}$ for the CUE and RM states indicated in the upper labels. The horizontal axis labels the 100 random states used in each one of the experiments. The rotation axes used are given in Eq.~\eqref{eq:axes}. The true values of $\omega$ are $0, 36, \ldots, 324$ degrees and are indicated by horizontal lines.} 
    \label{fig:results}
\end{figure*}

To perform the GPS-like measurement mentioned before, we probed each rotated random state $\ket{\psi_{\bm{\Omega}}}$ by projecting it onto a set of five coherent states, as described in detail in Ref.~\cite{Eriksson:2023aa}. {The explicit form of the FI matrix for this measurement is computed in Ref.~\cite{Goldberg:2021uw}.}  Effectively, our measurements produced the quantities
\begin{equation}
\mathcal{Q}_{\mathbf{n}} = | \langle \mathbf{n} |\psi_{\bm{\Omega}} \rangle|^2 = \frac{P_{\langle \mathbf{n} |\psi_{\bm{\Omega}} \rangle}}{P_{\langle \psi_{\bm{\Omega}} |\psi_{\bm{\Omega}} \rangle}} \frac{\eta_{\psi_{\bm{\Omega}}}}{\eta_{\mathbf{n}}} \, ,
\end{equation}
where $P_{\langle \mathbf{n} |\psi_{\bm{\Omega}} \rangle}$ was the relative power coupled into the final SMF when projecting the rotated random state $\ket{\psi_{\bm{\Omega}}}$ onto the coherent state $\ket{\mathbf{n}}$, and $P_{\langle \psi_{\bm{\Omega}} |\psi_{\bm{\Omega}} \rangle}$ was the power coupled when projecting the rotated random state onto itself.
The efficiencies $\eta_{\psi_{\bm{\Omega}}}$ and $\eta_{\mathbf{n}}$ are measured estimates of the state-dependent projection efficiencies: $\eta_{\psi_{\bm{\Omega}}}$ is the projection efficiency for the rotated random state measurement mask in question and similarly $\eta_{\mathbf{n}}$ is the efficiency of the CS projection.

The projection efficiencies were measured by first shaping the light field to the structure we want to project on in the generation, the MPLC was made to image the field structures, and the power of the field was measured both before the last SLM and after the last fiber. The efficiency was then estimated as the power after the fiber divided by the power before the measurement mask. The power measurements were normalized by dividing them with the power measured before the first spatial mode manipulation, as shown in Fig.~\ref{fig:setup}. This was done to minimize the effects of changes in laser power throughout the measurements.

As mentioned before, the probe states used in the experiment were both RM and CUE states. In our specific experimental implementation, where transverse spatial modes are used to encode different types of random states, the primary challenge in preparing, modulating, and measuring the states with minimal errors arises from the increasing dimensionality of the state space. This difficulty stems from the increasingly intricate transverse phase and amplitude structures, coupled with the limited resolution of our devices. Consequently, all types of states can be considered equally challenging to work with, regardless of their specific nature.

However, since the difficulty of generating states will vary across different systems, it is essential to weight the ease of generating each type of state against its metrological power. Generally, we anticipate that the more entangled a state is on average, such as CUE or N00N states, the harder it will be to generate them accurately. This suggests that RM states may serve as a valuable and powerful alternative.

It is important to stress that in our experimental setup  there is no need to perform any projection onto the symmetric subspace, as the states are directly generated in $\mathcal{H}_S$.  The concept of a symmetric subspace does not apply here, since our $\mathcal{H}{S}$ does not originate from $2S$ qubits.

 We performed the rotation measurement with 100 different random states of each type, with $\nu = 100$ repetitions of each relative-power measurement. The rotation axis $\mathbf{u}(\Theta,\Phi)$  was arbitrarily chosen for each value of $S$:
\begin{equation}
\mathbf{u} (\Theta, \Phi) = \left \{ 
\begin{array}{ll}
\Theta = 34.9674^\circ  ,  \Phi = 296.4448^\circ & \quad S=2, \\
& \\
\Theta = 157.948^\circ, \Phi = 137.361^\circ & \quad S= 3 \, .
\end{array}
\right .
\label{eq:axes}
\end{equation}
Three different kinds of generalized maximum likelihood estimators were formed from this data set to verify the excellent rotational sensitivity of the fiducial states, following standard procedures~\cite{Hradil:2006aa,Rehacek:2008aa}. The estimator of the rotated angle $\omega$ reads
\begin{equation}
\hat{\omega} = \arg \max_\omega \sum_{j=1}^{5} \mathfrak{q}_j \log\frac{\mathcal{Q}_j}{\sum_{j^{\prime}} \mathcal{Q}_{j^{\prime}}}\,,
\end{equation}
where $\mathfrak{q}_{j}$ is the measured value of $\mathcal{Q}_{j}= | \langle \mathbf{n}_{j} |\psi_{\bm{\Omega}} \rangle|^2$. This is equivalent to a quantum measurement where there are five meaningful outcomes, the measurement is repeated $\nu$  times, and the frequency with which each outcome occurs is proportional to $\mathfrak{q}_{j}$. The results appear in Fig.~\ref{fig:results}. The estimated values of $\omega$ show small fluctuations around the true values with very small typical standard deviations and negligible biases. The three cases look very similar.

\begin{figure}[t]
\includegraphics[width=\columnwidth]{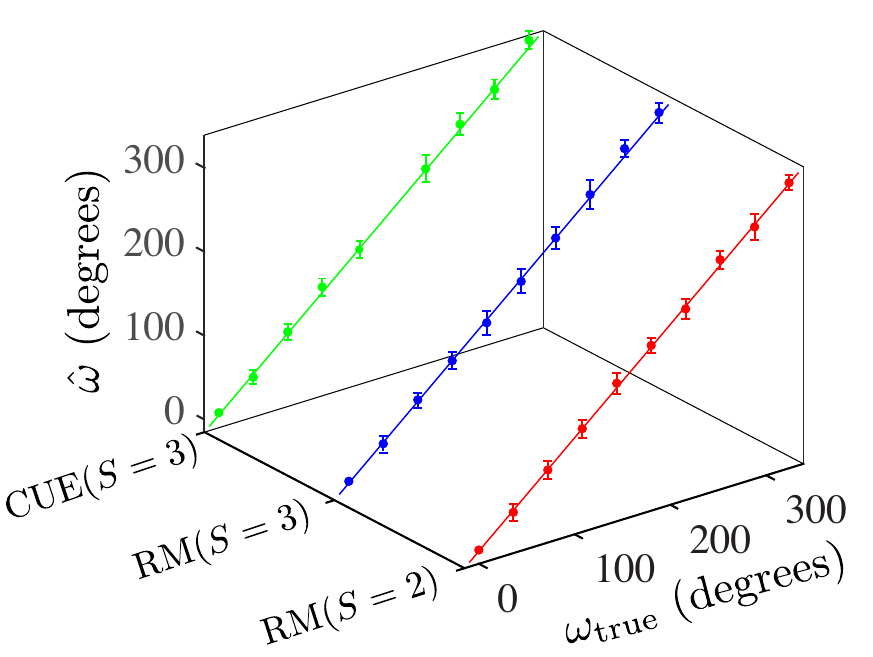}
\caption{Statistics of the measurement for the states considered before in Fig.~\ref{fig:results}. The error bars represent standard deviations. {The estimation procedure is local: the same states can be used for estimating a wide range of parameter values, but prior information about those parameters is essential to obtain the sensitivity limits dictated by local estimation theory.}}
\label{fig:estaxis}
\end{figure}

To check the quality of our estimators, in Fig.~\ref{fig:estaxis} we have plotted the true values of $\omega$ versus the estimated ones. For each estimated point, we have included error bars determined by the corresponding standard deviation. The agreement emphasizes the excellent performance of the method. {Moreover, Fig.~\ref{fig:estaxis} corroborates a unique property of random states: due to their typicality~\cite{Garnerone:2010aa}, they can unambigously estimate any rotation angle within a $2\pi$ range. This is in sharp contrast with other states such as NOON or  of Quantumness that can distinguish angles within a range $\pi/(2S)$~\cite{Bjork:2015vv}.}

\begin{figure}[t]
\includegraphics[width=0.80\columnwidth]{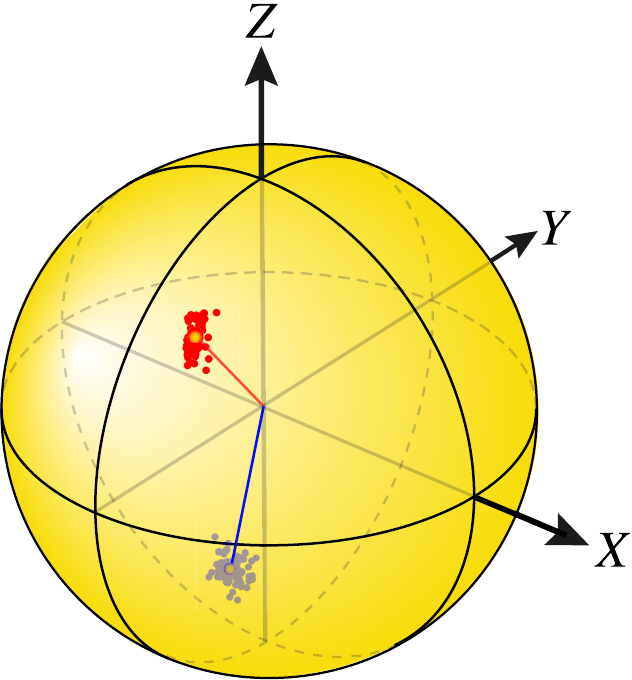}
\caption{Estimated rotation axes for RM with $S=2$ (red points) and $S=3$ (blue points) using the data from Fig.~\ref{fig:results}.}
\label{fig:sphereaxis}
\end{figure}

Turning the assumptions around, we can take now the set of rotation angles as fixed, while assuming no prior knowledge of the rotation axis. The orientation of this axis  can be estimated globally as follows
\begin{equation}
\hat{\mathbf{u}} = \arg \max_{\mathbf{u}} \sum_k\sum_j \mathfrak{q}_{jk}\log\frac{\mathcal{Q}_{jk}}{\sum_{k^{\prime}}\sum_{j^{\prime}} \mathcal{Q}_{k^{\prime}j^{\prime}}} \, ,
\end{equation}
where now $\mathcal{Q}_{jk}=|\langle \mathbf{n}_j|\psi_{k,\Omega}\rangle|^2$. This time, there is a bigger set of outcomes: each of the five quantum measurement outcomes for each of the rotation angles. Using this on the ensemble of $100$ input RM states gives  the statistics of inferred axis directions 
: 
\begin{equation}
\hat{\mathbf{u}} (\hat{\Theta}, \hat{\Phi}) = \left \{ 
\begin{array}{ll}
\hat{\Theta} = 35.31^{\circ} \pm 4.0^{\circ}  ,  \hat{\Phi} = 295.8^\circ \pm 2.9^{\circ}& \quad S=2, \\
& \\
\hat{\Theta} = 138.4^{\circ}\pm 2.9^{\circ}, \hat{\Phi} = 157.2^{\circ}\pm 3.9^{\circ} & \quad S= 3 \, .
\end{array}
\right .
\label{eq:axesest}
\end{equation}
All the target true values are comfortably accommodated within those uncertainty regions. This estimation is nicely visualized on the sphere $\mathcal{S}_2$, as sketched in Fig.~\ref{fig:sphereaxis}. The results for CUE random states are similar, although a bit less reliable due to experimental difficulties in getting the corresponding data. 

These values can be compared with the best possible results respectively for the variances of $\hat{\Theta}$ and $\hat{\Phi}$ for a single measurement with a CS, which are (albeit for different orientations of CS)
\begin{equation} 
{\Var_{\mathrm{CS}} (\hat{\Theta}) = \Var_{\mathrm{CS}} (\hat{\Phi}) = \frac{1}{2S} \, }
\end{equation} 
and the variances with Kings of Quantumness
\begin{equation}
\begin{aligned} 
{\Var_{\mathrm{Kings}} ( \hat{\Theta})} &  =  
\frac{3}{16 S(S+1) \sin^2 (\omega/2)} \, ,  \\
{\Var_{\mathrm{Kings}} (\hat{\Phi})} &  = {\frac{3}{16 S(S+1) \sin^2(\omega/2) \sin^2\Theta} \, .}
\end{aligned}
\end{equation}

{When performing such a multiparameter estimation, the achievable uncertainties tend to depend on the actual values of the parameters. This dependence can be removed by choosing an appropriate weight matrix for combining the covariances of the different parameters: using the  metric tensor for the group as the weight matrix allows all of the results to be independent from the parameter values and the chosen parametrization\cite{Goldberg:2021vj}. As in Ref.~\cite{Eriksson:2023aa}, such a figure of merit when estimating the angular coordinates of a unit vector is $\Delta^2 \hat{\Theta}+\sin^2\hat{\Theta} \Delta^2 \hat{\Phi}$, here 0.00488634 and 0.00257369, as compared to the best possible values with a single coherent state of}
\begin{equation}
{\frac{1}{2S}\frac{2 \cos \omega+ \csc^2(\omega/2)+2}{2 \sqrt{2} \sin \omega +\cos \omega +3}> \frac{0.3758}{2S} \, . }
\end{equation}

Finally,  our detection scheme projecting onto five coherent states has a redundancy that  makes possible to estimate simultaneously the axis of rotation and the angle of rotation. This implies maximizing the objective function
\begin{equation}
\{\hat{\omega},\hat{\mathbf{u}} \}=\arg\max_{\{\omega, \mathbf{u}\}}  \sum_{j=1}^5 \mathfrak{q}_j \log\frac{\mathcal{Q}_j}{\sum_{j^{\prime}}\mathcal{Q}_{j^{\prime}}} \, ,
\end{equation}
which requires global optimization tools~\cite{GLOBAL:1995aa}. Again, we are using a classical system to simulate a quantum measurement by taking the frequency of quantum measurement outcomes to be proportional to the relative power received for each of the five settings, then finding the variances of the maximum likelihood estimators that we take to be distributed according to the underlying probability distribution for the true parameters.

\begin{figure}[t]
\includegraphics[width=\columnwidth]{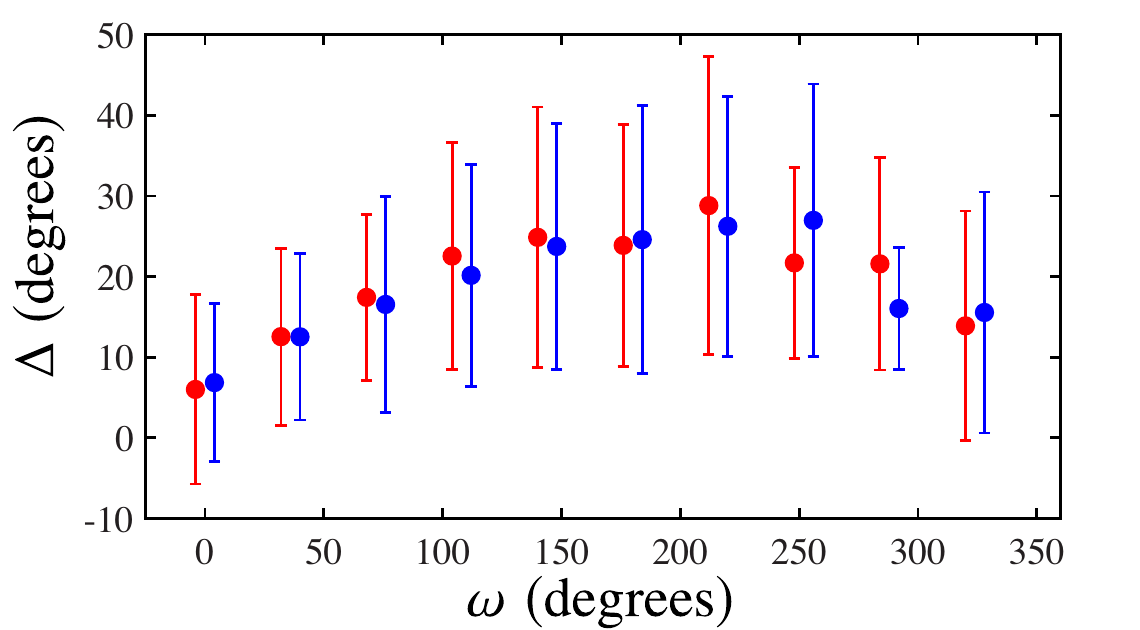}
\caption{Deviation $\Delta$ between the true rotation and the estimated rotation for RM states with $S=2$ (red) and $S=3$ (blue), determined by the angle of rotation one would require to convert between the true and estimated rotation, as a function of the angle. The large error bars on the deviations make these results compatible with both quantum advantages (small deviations) and lack of demonstrated quantum advantages (large deviations); in all cases, the measurement method is demonstrated to be viable for determining all three parameters of a rotation.}
\label{fig:3Dest}
\end{figure}

Errors of the inferred three parameters of the  rotations can be quantified by considering the deviation $\Delta$ between the true rotation and the estimated rotation, determined by the angle of rotation one would require to convert between the true and estimated rotation, as $\Delta=\arg R(-\bm{\Omega})R(\hat{\bm{\Omega}})$, which can be found from the Hilbert-Schmidt norm as $\cos\Delta S+\sin\Delta S/\tan(\Delta/2)=\Tr[R(-\bm{\Omega})R(\hat{\bm{\Omega}})]$~\cite{Eriksson:2023aa}. This way, we find the average deviations
\begin{equation}
\Delta = \left \{ 
\begin{array}{ll} 
19.3^{\circ}\pm 15.2^{\circ} & \qquad S=2 \, , \\
& \\
18. 9^{\circ}\pm 15.1^{\circ} & \qquad S=3 \, .
\end{array}
\right .
\end{equation} 
The results are plotted for RM states in Fig.~\ref{fig:3Dest}. {These can be compared with the best possible results for measurement with a single  CS as a probe state, which can never be used to estimate all three parameters of a rotation due to CS only depending on two coordinates that define the state's spin. Comparisons of schemes that split all of the trials into different subsets that each involve the same respective probe state can be found in the recent Ref.~\cite{Ferretti:2024aa}. This allows us to conclude that the high rotational sensitivity of the RM states has been experimentally verified.

{Although in our experiment the rotations have been artificially encoded using an MPLC system, the coherent-state projections used for the detection offer a range of advantages, including compactness, versatility, high efficiency, flexibility,  and real-time adaptability.  As already demonstrated in Ref.~\cite{Eriksson:2023aa}, this scheme is  broad enough to apply to any other rotation-sensing application.}

\section{Concluding remarks}
\label{sec:conc}

We have explored random Majorana constellations that arise as sets of points uniformly distributed on the sphere $\mathcal{S}_{2}$.  The concept of state multipoles, intimately linked with the inherent SU(2) symmetry, has served as our main diagnostic tool to capture the amazing properties of these states. Additionally, these multipoles are sensible and experimentally-realizable quantities.

The family of symmetric states contains many metrologically useful states, including GHZ, NOON, and Dicke states, among others. However, all of them are extremely fragile resources. In contradistinction, random Majorana states, from their very same definition, are robust against imperfections such as dephasing and particle losses.  

We have experimentally demonstrated the  usefulness of these random constellations in protocols of quantum metrology. Apart from their incontestable geometrical beauty, there surely is plenty of room for the application of these states in a variety of physical contexts.

\bigskip

\section*{Acknowledgments} 
We are indebted to M.~Grassl, H.~Ferretti, A.~Steinberg, and K.~\.{Z}yczkowski for discussions. This work was supported by the European Union's Horizon 2020 research and innovation programme under the QuantERA programme through the project ApresSF. AZG  acknowledges that the NRC headquarters is located on the traditional unceded territory of the Algonquin Anishinaabe and Mohawk people and support from NRC's Quantum Sensors Challenge Program and an NSERC Postdoctoral Fellowship. ABK acknowledges the Mexican CONAHCyT (Grant CBF2023-2024-50). MH acknowledges the Doctoral School of Tampere University, the Magnus Ehrnrooth Foundation, and the Emil Aaltonen Foundation. RF acknowledges support from the Academy of Finland through the Academy Research Fellowship (Decision 332399). ME, MH, RF acknowledge the support of the Academy of Finland through the Photonics Research and Innovation Flagship (PREIN-decision 320165). LLSS acknowledges support from Ministerio de Ciencia e Innovaci\'on (Grant  PID2021-127781NB-I00).  


%

\end{document}